\documentclass[letterpaper,twocolumn,preprintnumbers,amsmath,amssymb,amsfonts,floatfix,nofootinbib]{revtex4}

\usepackage{graphicx}
\usepackage{amssymb}
\usepackage{amsmath}
\usepackage{hypernat}
\usepackage{color}
\newif\ifcomment
\newif\ifprint
\printtrue

\ifprint
 \usepackage[linkbordercolor={0 0 .8},%
             citebordercolor={.8 0 0},%
             urlbordercolor={.8 0 .8},%
             raiselinks=true,%
             pdfborder={0 0 1 [3]}]{hyperref}
\else
 \usepackage[pdftex,backref,pagebackref,colorlinks,
             linkcolor=blue,citecolor=blue,anchorcolor=blue,
             pagecolor=blue,urlcolor=red]{hyperref}
\fi
\pdfoutput=1        
\pdfcatalog{/PageMode(/UseOutlines)} 

\newcommand {\snn}         {\ensuremath{\sqrt{s_{\scriptscriptstyle{{\rm NN}}}}}}
\newcommand {\signn}       {\ensuremath{\sigma_{\scriptscriptstyle{{\rm NN}}}}}
\newcommand {\ep}          {\mbox{$\epsilon_{\rm part}$}}
\newcommand {\erp}         {\mbox{$\epsilon_{\rm RP}$}}
\newcommand {\Ncoll}       {\ensuremath{N_{\rm coll}}}
\newcommand {\Npart}       {\ensuremath{N_{\rm part}}}
\newcommand {\arxiv}[1]    {\href{http://www.arxiv.org/abs/#1}{\mbox{arXiv:#1}}}
\newcommand {\hrefurl}[1]  {\href{#1}{\url{#1}}}
\newcommand {\Ref}[1]      {Ref.~\cite{#1}}

\newcommand {\Fig}[1]      {Fig.~\ref{#1}}
\newcommand {\hide}[1]     {\color{white}#1\color{black}}
\newcommand {\lmyangle}    {\ensuremath{\{}}
\newcommand {\rmyangle}    {\ensuremath{\}}}

\begin{document}

\title{The PHOBOS Glauber Monte Carlo}
\author{B.Alver$^1$, M.Baker$^2$, C.Loizides$^1$, P.Steinberg$^2$}
\affiliation{
$^1$Massachusetts Institute of Technology, Cambridge, MA 02139, USA\\
$^2$Brookhaven National Laboratory, Upton, NY 11973, USA
}

\begin{abstract}\noindent
``Glauber'' models are used to calculate geometric quantities in the
initial state of heavy ion collisions, such as impact parameter,
number of participating nucleons and initial eccentricity. The four
RHIC experiments have different methods for Glauber Model calculations, 
leading to similar results for various geometric
observables. In this document, we describe an implementation of the
Monte Carlo based Glauber Model calculation used by the PHOBOS
experiment. The assumptions that go in the calculation are
described. A user's guide is provided for running various calculations.
\vspace{3mm}
\noindent 
\end{abstract}
\maketitle

\section{Introduction}\label{sec:intro}
In heavy-ion collisions, initial geometric quantities such as impact
parameter and shape of the collision region cannot be directly
determined experimentally. However, it is possible to relate the
number of observed particles and number of spectator neutrons to the
centrality of the collision. Using the percentile centrality of a
collision, the initial geometric configuration can be estimated with
models of the contents of a typical nucleus. 

These models fall in two main classes. (For a recent review, see
\Ref{glaubreview}.) In the so called ``optical'' Glauber
calculations, a smooth matter density is assumed, typically described
by a Fermi distribution in the radial direction and uniform over solid
angle. In the Monte Carlo based models, individual nucleons are
stochastically distributed event-by-event and collision properties are
calculated by averaging over multiple events. As discussed in
\Ref{glaubreview} and \Ref{eccentricity}, these two type of models
lead to mostly similar results for simple quantities such as the number of
participating nucleons~($\Npart$) and impact parameters~($b$), but
give different results in quantities where event-by-event fluctuations
are significant, such as participant frame eccentricity~($\ep$).

In this paper, we discuss in detail the Monte Carlo Glauber
calculation implemented by PHOBOS. In section~\ref{sec:method}, the
method is outlined and the assumptions that go into the calculation
are introduced. In section~\ref{sec:howto}, we discuss the
implementation and the tutorial functions provided.

\section{The Model}\label{sec:method}
The Monte Carlo Glauber Model calculation is performed in two
steps. At first, the nucleon positions in each nucleus are stochastically
determined. Then, the two nuclei are ``collided'', assuming the
nucleons travel in a straight line along the beam axis (eikonal approximation) 
such that nucleons are tagged as wounded~(participating) or spectator.

\subsection{Makeup of Nuclei}\label{sec:nucleus}
The position of each nucleon in the nucleus is determined according to
a probability density function. In a quantum mechanical picture, 
the probability density function can be thought of as the single-particle
probability density and the position as the result of a position
measurement. In the determination of the nucleon positions in a given
nucleus, it is possible to require a minimum inter-nucleon 
separation~($d_{\rm min}$) between the centers of the nucleons.

The probability distribution is typically taken to be uniform in
azimuthal and polar angles. 
The radial probability function is modeled from nuclear charge densities 
extracted in low-energy electron scattering experiments~\cite{atomicdata}. 
The nuclear charge density is usually parameterized by a Fermi
distribution with three parameters:
\begin{equation}
  \rho(r)=\rho_0 \frac{1+w(r/R)^2}{1+exp(\frac{r-R}{a})}\,,
\end{equation} 
where $\rho_0$ is the nucleon density, $R$ is the nuclear radius, $a$
is the skin depth and $w$ corresponds to deviations from a spherical
shape. The overall normalization ($\rho_0$) is not relevant for this
calculation. Values of the other parameters used for different nuclei
are listed in Table~\ref{tab:awR}.

Two exceptions are the deuteron (${}^{2}$H) and sulfur (${}^{32}$S)
nuclei. For sulfur, a three parameter Gaussian form is used:
\begin{equation}
  \rho(r)=\rho_0 \frac{1+w(r/R)^2}{1+exp(\frac{r^2-R^2}{a^2})}\,.
\end{equation} 
The values of $R$, $a$ and $w$ for sulfur are also given in
Table~\ref{tab:awR}. For deuteron, three options are supported:
\begin{enumerate}
\item The three parameter Fermi distribution can be used, with the
 values given in Table~\ref{tab:awR}.
\item The Hulth\'en form can be used:
\begin{equation}
  \rho(r)=\rho_0 \left(\frac{e^{-ar}+e^{-br}}{r}\right)^2,
\end{equation} 
where $a=0.457$~fm$^{-1}$ and $b=2.35$~fm$^{-1}$ \cite{Hulthen, deuteronpars}. 
\item The proton can be randomly placed using the Hulth\'en form given
above and the neutron can be placed opposite to it. 
\end{enumerate}
It should be noted that the $3^{\text{rd}}$ option was used
in PHOBOS analyses.

\begin{table}[t]
\begin{center}
  \begin{tabular}{|c|c|c|c|}
    \hline
    Nucleus    & R~[fm] &  a~[fm]  & w~[fm]    \\
    \hline
    ${}^{2}$H    & 0.01\hide{0}  & 0.5882        & \hide{-}0\hide{.0000}       \\
    ${}^{16}$O   & 2.608         & 0.513\hide{0} & -0.51\hide{00}              \\
    ${}^{28}$Si  & 3.34\hide{0}  & 0.580\hide{0} & -0.233\hide{0}              \\
    ${}^{32}$S   & 2.54\hide{0}  & 2.191\hide{0} & \hide{-}0.16\hide{00}       \\
    ${}^{40}$Ca  & 3.766         & 0.586\hide{0} & -0.161\hide{0}              \\
    ${}^{58}$Ni  & 4.309         & 0.517\hide{0} & -0.1308                     \\
    ${}^{62}$Cu  & 4.2\hide{00}  & 0.596\hide{0} & \hide{-}0\hide{.0000}       \\
    ${}^{186}$W  & 6.58\hide{0}  & 0.480\hide{0} & \hide{-}0\hide{.0000}       \\
    ${}^{197}$Au & 6.38\hide{0}  & 0.535\hide{0} & \hide{-}0\hide{.0000}       \\
    ${}^{207}$Pb\footnote{These values are also used for ${}^{208}$Pb for which Fermi 
                          parameters are not available. It has been noted that Bessel-Fourier 
                          coefficients for the two nuclei are similar~\cite{atomicdata}.} 
                 & 6.62\hide{0}  & 0.546\hide{0} & \hide{-}0\hide{.0000}       \\
    ${}^{238}$U  & 6.81\hide{0}  & 0.6\hide{000} & \hide{-}0\hide{.0000}       \\
    \hline
  \end{tabular}
  \caption{\label{tab:awR}\protect Nuclear charge density parameters for different nuclei, 
           taken from \Ref{atomicdata}.}
\end{center}  
\end{table}

\begin{figure}
\begin{center}
   \includegraphics[width=80mm]{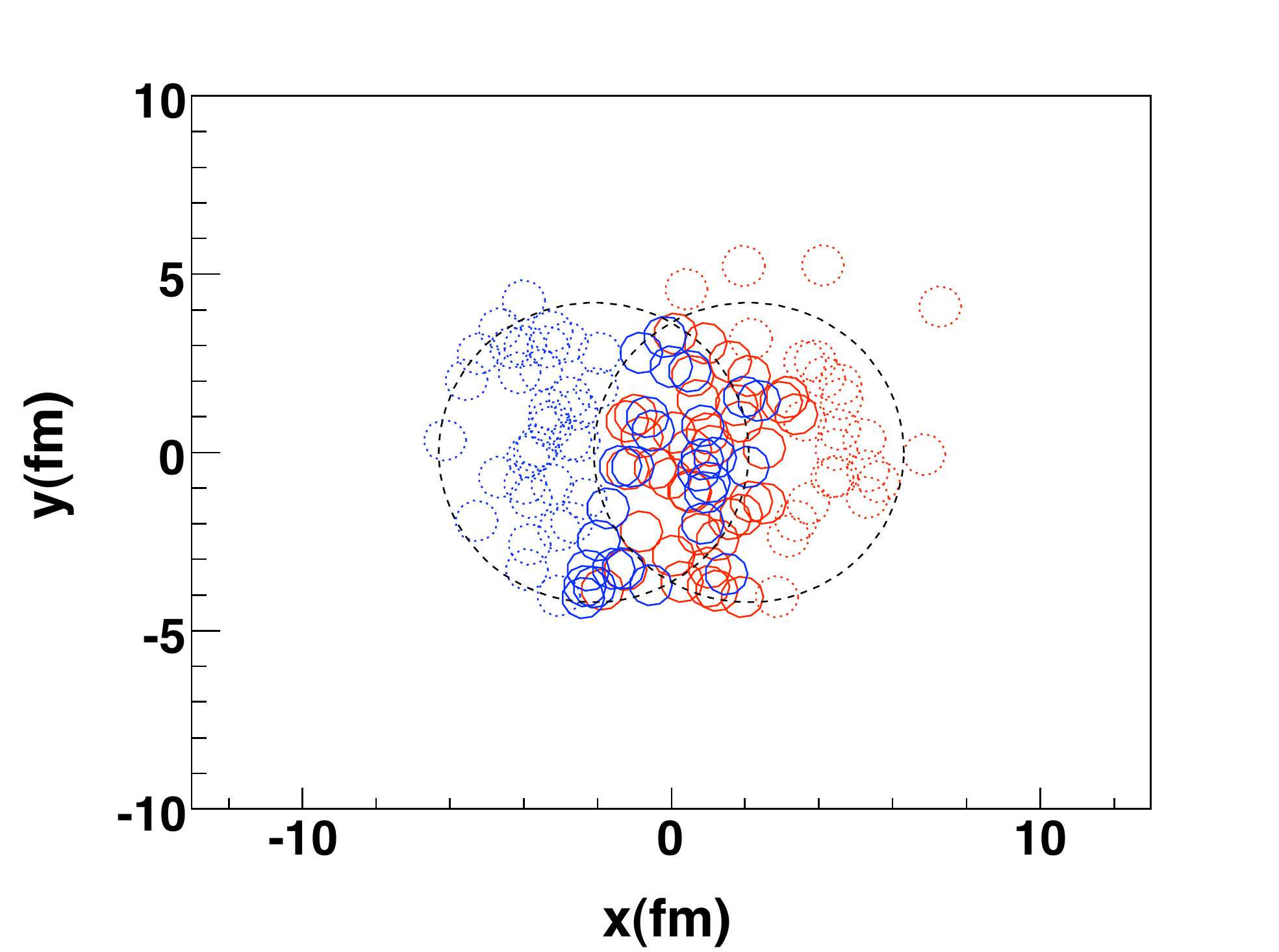}
   \includegraphics[width=80mm]{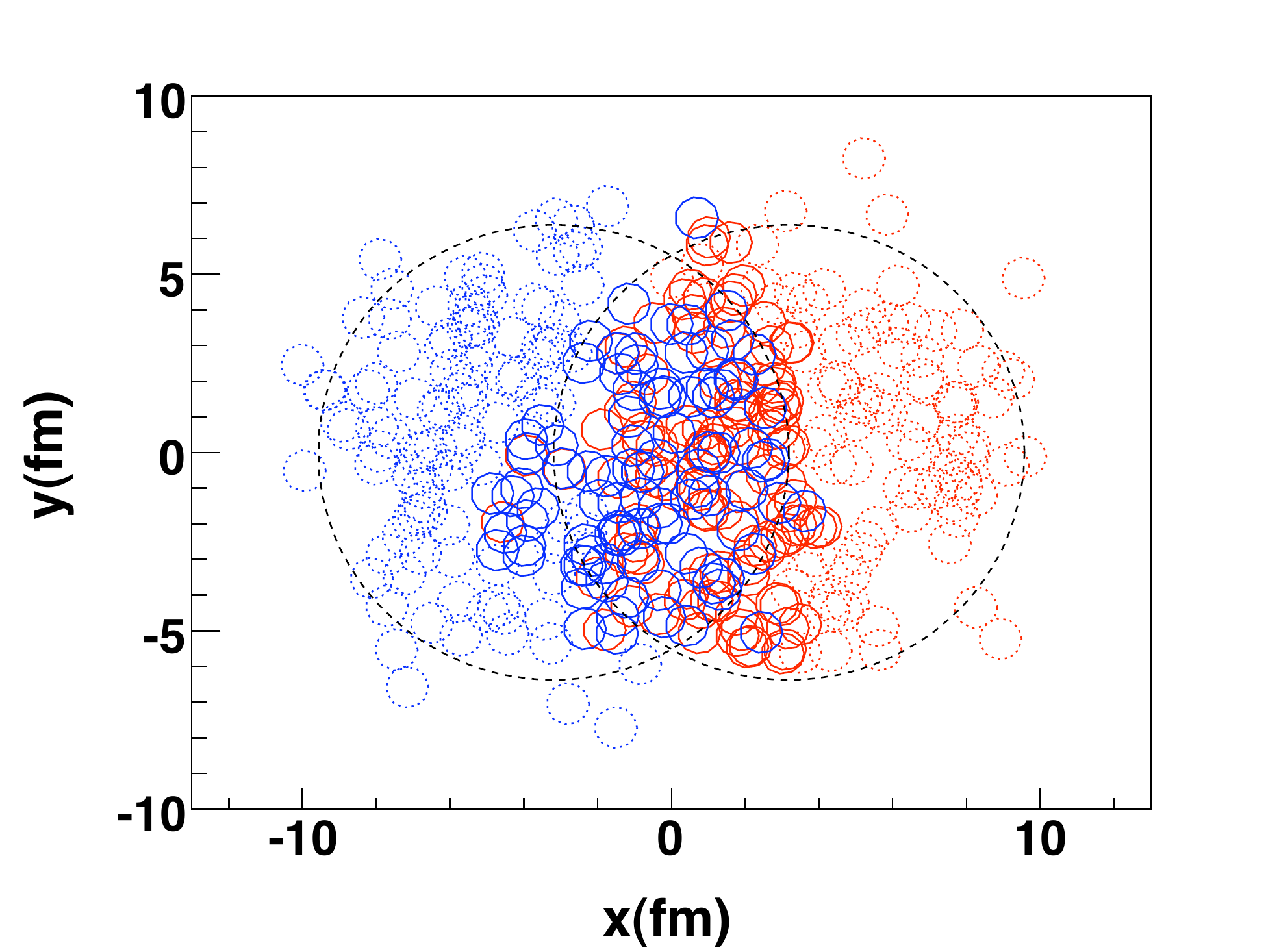}
   \includegraphics[width=80mm]{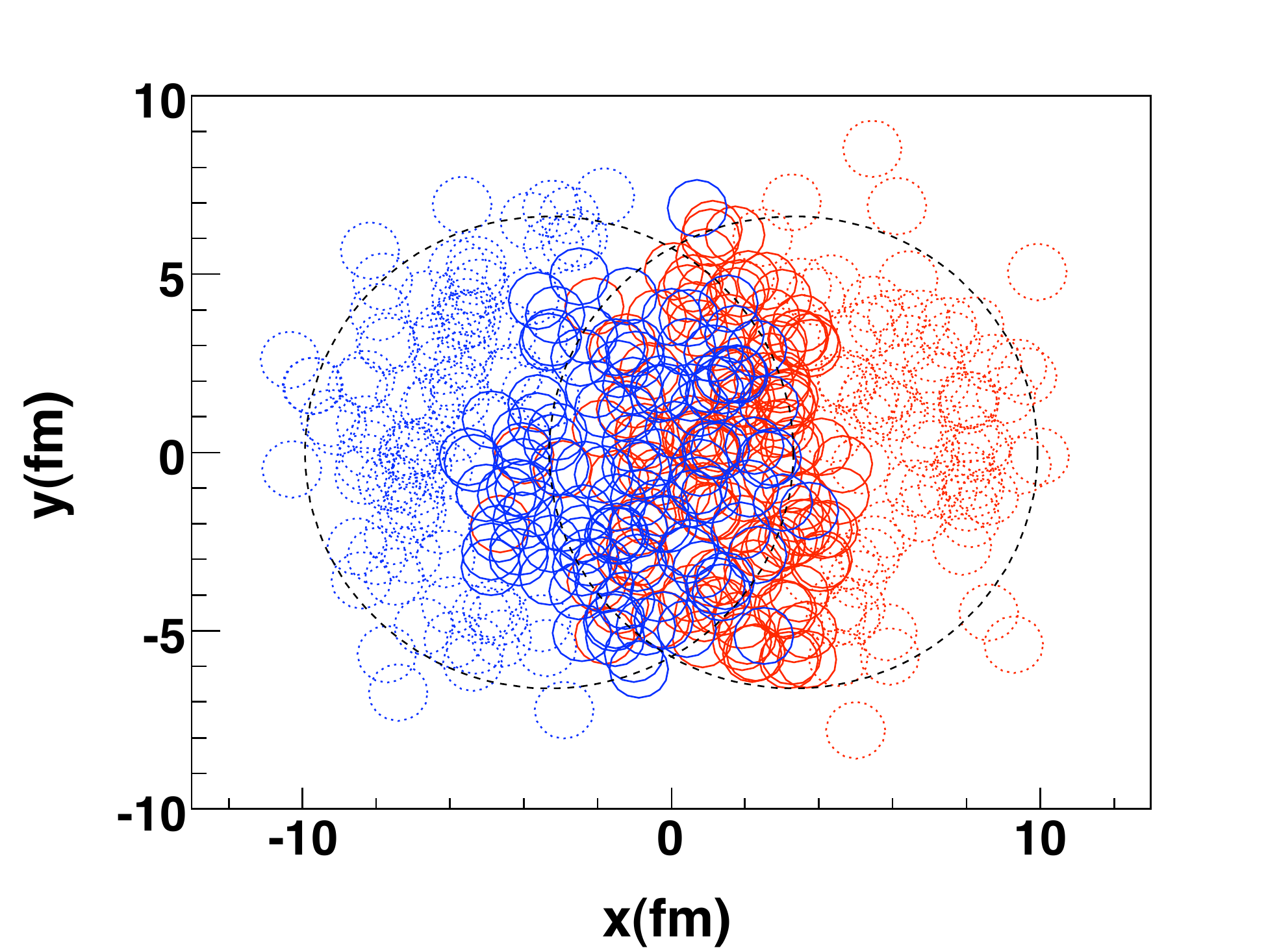}
   \caption{\label{fig:testplots}Typical events for Cu+Cu~(top panel), Au+Au~(middle panel), 
            and Pb+Pb~(lower panel) collisions, the first two performed at RHIC energies 
            and the latter at the LHC. Wounded nucleons~(participants) are indicated as 
            solid circles, while spectators are dotted circles.}
\end{center}
\end{figure}

\subsection{Collision Process}\label{sec:coll}
The impact parameter of the collision is chosen randomly from a
distribution ${\rm d}N/{\rm d}b \propto b$ up to some large maximum
$b_{\rm max}$ with \mbox{$b_{\rm max}\simeq20\,$fm$>2R_{A}$}. 
The centers of the nuclei are calculated and shifted to $(-b/2,0,0)$ 
and $(b/2,0,0)$~\footnote{Throughout the paper, the reaction 
plane, defined by the impact parameter and the beam direction, is given
by the $x$- and $z$-axes, while the transverse plane is given by the 
$x$- and $y$-axes.}.
It is assumed that the nucleons move
along a straight trajectory along the beam axis.~(The longitudinal
coordinate does not play a role in the calculation.)

The inelastic nucleon-nucleon cross section ($\signn$), which is
only a function of the collision energy is extracted from p+p
collisions. At the top RHIC energy of $\snn=200$~GeV, $\signn=42$~mb, 
while at the LHC it is expected to be around $\signn=72$~mb~(with large 
uncertainty from the unknown elastic cross section). 
The ``ball diameter'' is defined as:
\begin{equation}
  D = \sqrt{\sigma_{NN}/\pi}.
\end{equation} 
Two nucleons from different nuclei are assumed to collide if their
relative transverse distance is less than the ball diameter. 
If no such nucleon--nucleon collision is registered for any pair of nucleons,
then no nucleus--nucleus collision occurred. Counters for determination of the 
total~(geometric) cross section are updated accordingly.

\section{Users' Guide}\label{sec:howto}
The PHOBOS Glauber MC code works within the ROOT framework (ROOT 4.00/08 or
higher~\cite{root}). The code is contained in the macro 
{\tt runglauber\_vX.Y.C}~\cite{glaucode}~(Latest version is 1.0.).
Three classes, {\tt TGlauNucleon}, {\tt TGlauNucleus} and {\tt TGlauberMC} 
and two example functions {\tt runAndSaveNtuple()} and {\tt runAndSaveNucleons()}
are defined in the provided macro. 
While the functionality is essentially complete 
for known applications of the Glauber approach, users are encouraged to write their
own functions to access results of the Glauber simulation or to modify the code:

\begin{itemize}
\item{\tt TGlauNucleon} is used to store information about a single nucleon. 
The stored quantities are the position of the nucleon, the number of binary collisions 
that the nucleon has had and which nucleus the nucleon is in, ``A'' or ``B''. 
For every simulated event, the user can obtain an array containing all nucleons~(via {\tt 
TGlauberMC::GetNucleons()}).
\item{\tt TGlauNucleus} is used to generate and store information about a 
single nucleus. The user is not expected to interact with this class.
\item{\tt TGlauberMC} is the main steering class used to generate events and 
calculate event-by-event quantities such as the number of participating nucleons. 
\end{itemize}

The steering class {\tt TGlauberMC} has one constructor
\begin{verbatim}
TGlauberMC::TGlauberMC(Text_t* NA, 
                       Text_t* NB, 
                       Double_t xsect)
\end{verbatim}
where {\tt NA} and {\tt NB} are the names of the colliding nuclei
and {\tt xsect} is the nucleon-nucleon cross section given in mb. The
defined nuclei names are: ``p'', ``d'', ``dhh'', ``dh'', ``O'',
``Si'', ``S'', ``Ca'', ``Cu'', ``W'', ``Pb'', ``Au'', ``Ni'' and
``U'' (see Table~\ref{tab:awR}). For deuteron, the names ``d'', ``dhh'' and 
``dh'' correspond to the three options described in section~\ref{sec:nucleus} 
respectively. Units are generally given in fm for distances, while in mb 
for cross sections.

\subsection{Running the Code}
To generate Au+Au collisions at $\sqrt{s_{_{\it NN}}}=200$~GeV~($\snn=42$~mb) one would 
construct a {\tt TGlauberMC} object by issuing the commands:
\begin{verbatim}
root [0] .L runglauber_X.Y.C+
root [1] TGlauberMC glauber("Au","Au",42);
\end{verbatim}
where the first ROOT command compiles, links and loads the compiled macro~\footnote{Note
that you must replace X.Y with the current version number of the code, for
example~1.1.}
including the Glauber code as explained in chapter~2 of the ROOT users' guide. 

Events can be generated interactively using the two functions
\begin{itemize} 
\item{\tt TGlauberMC::NextEvent(Double\_t bgen)}, which is used to run an event 
at a specified impact parameter, or over a range of impact parameters~(if {\tt bgen=-1}, 
the default value) as described in section~\ref{sec:coll}).
\item{\tt TGlauberMC::Run(Int\_t nevents)} which is used to run a large event sample 
by invoking {\tt NextEvent} many times.
\end{itemize}

\begin{figure}[t!f]
\begin{center}
  \includegraphics[width=80mm]{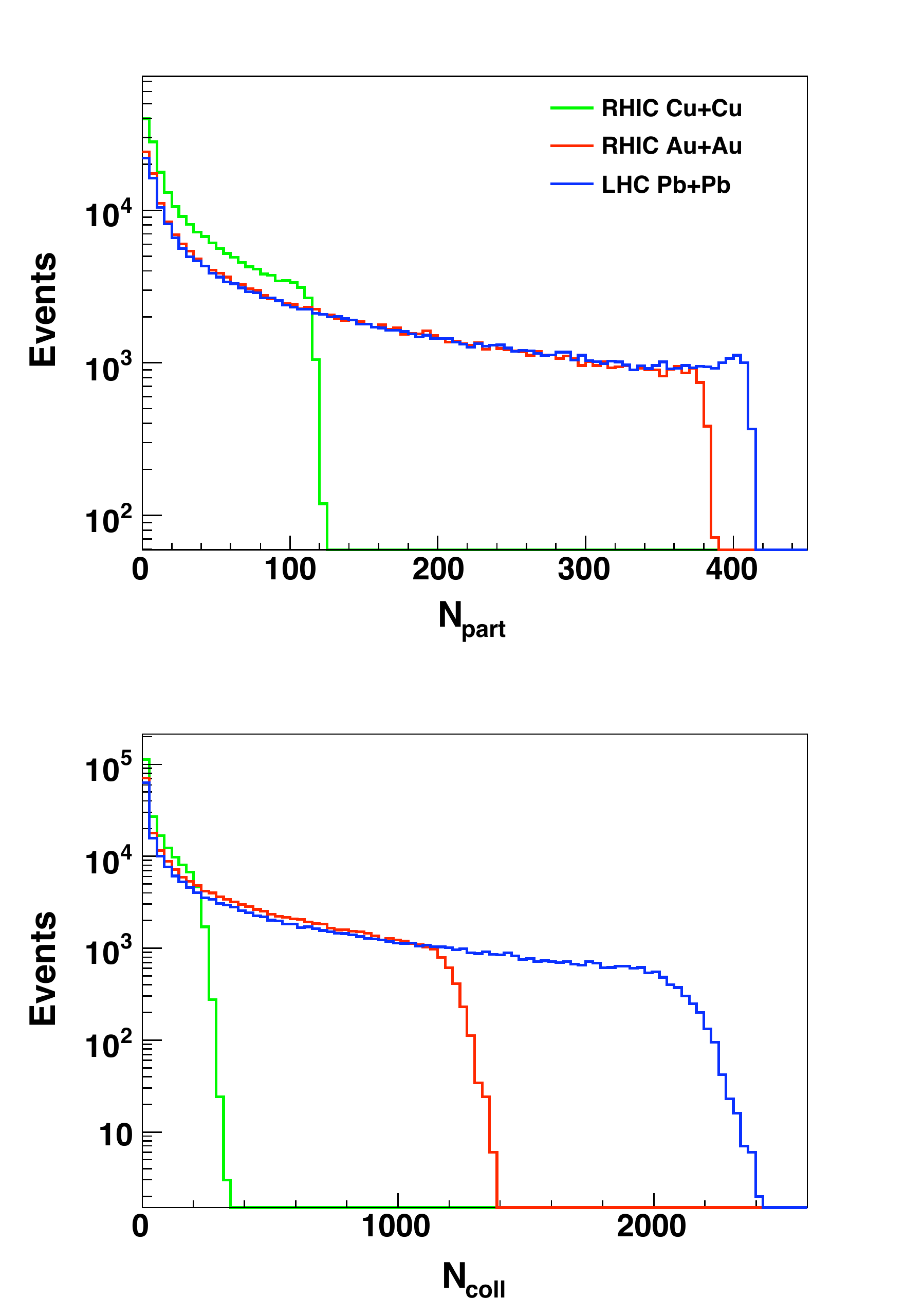}
  \caption{\label{fig:npcplots}Distributions of $\Npart$ and $\Ncoll$ for 10k events for 
           Cu+Cu and Au+Au at RHIC, and Pb+Pb at the LHC.}
\end{center}
\end{figure}

Other important public member functions are:
\begin{itemize}
\item {\tt TGlauberMC::SetMinDistance(Double\_t d)}, which is used to set
minimum nucleon seperation within a nucleus, $d_{\rm min}$ (default is $0.4$~fm)
\item {\tt TGlauberMC::SetBmin(Double\_t bmin)} and {\tt TGlauberMC::SetBmax(Double\_t bmax)}, 
which can be used to set the range of impact parameter values generated in {\tt Run()}.
\item {\tt TGlauberMC::GetTotXSect()}
which returns the total nucleus-nucleus cross section, calculated when
the function {\tt Run()} is called.
\item {\tt TGlauberMC::Draw()} which draws the current event in the current pad.
\end{itemize}

\subsection{Example functions}
Two example functions are provided to demonstrate how to run the model. 

{\tt runAndSaveNtuple()} generates a number of Monte Carlo events and
saves some event-by-event quantities. It takes as parameters, the
number of events to be generated, the collision system, the
nucleon-nucleon cross section, the minimum separation distance 
and the output file name. It creates and stores an ntuple in the 
output file with the following event-by-event quantities:
\begin{itemize}
\item{\tt Npart}:       Number of participating nucleons.
\item{\tt Ncoll}:       Number of binary collisions.
\item{\tt B}:           Generated impact parameter.
\item{\tt MeanX}:       Mean of $x$ for wounded nucleons, $\lmyangle x \rmyangle$.
\item{\tt MeanY}:       Mean of $y$ for wounded nucleons, $\lmyangle y \rmyangle$.
\item{\tt MeanX2}:      Mean of $x^2$ for wounded nucleons, $\lmyangle x^2 \rmyangle$.
\item{\tt MeanY2}:      Mean of $y^2$ for wounded nucleons, $\lmyangle y^2 \rmyangle$.
\item{\tt MeanXY}:      Mean of $xy$ for wounded nucleons, $\lmyangle xy \rmyangle$.
\item{\tt VarX}:        Variance of $x$ for wounded nucleons, $\sigma_x^2$.
\item{\tt VarY}:        Variance of $y$ for wounded nucleons, $\sigma_y^2$.
\item{\tt VarXY}:       Covariance of $x$ and $y$ for wounded nucleons, 
                        $\sigma_{xy} \equiv \lmyangle xy \rmyangle - \lmyangle x \rmyangle \lmyangle y \rmyangle$.
\item{\tt MeanXSystem}: Mean of $x$ for all nucleons.
\item{\tt MeanYSystem}: Mean of $y$ for all nucleons.
\item{\tt MeanXA}:      Mean of $x$ for nucleons in nucleus A.
\item{\tt MeanYA}:      Mean of $y$ for nucleons in nucleus A.
\item{\tt MeanXB}:      Mean of $x$ for nucleons in nucleus B.
\item{\tt MeanYB}:      Mean of $y$ for nucleons in nucleus B.
\end{itemize}
It is important to note that for each of these event-by-event quantities a ``getter'' function
is implemented providing the users the option to write their own event loop~(using {\tt 
TGlauberMC::NextEvent()}.

The function {\tt runAndSaveNucleons()} generates a number of Monte Carlo
events and saves an array of {\tt TGlauNucleon} objects for each event.  
It is also possible to use this function to print out the values stored
in the nucleons by setting the verbosity parameter.
The function takes as parameters the number of events to be generated, 
the collision system, the nucleon-nucleon cross section, the minimum separation 
distance, the verbosity flag and the output file name.

\begin{figure}[t!f]
\begin{center}
  \includegraphics[width=80mm]{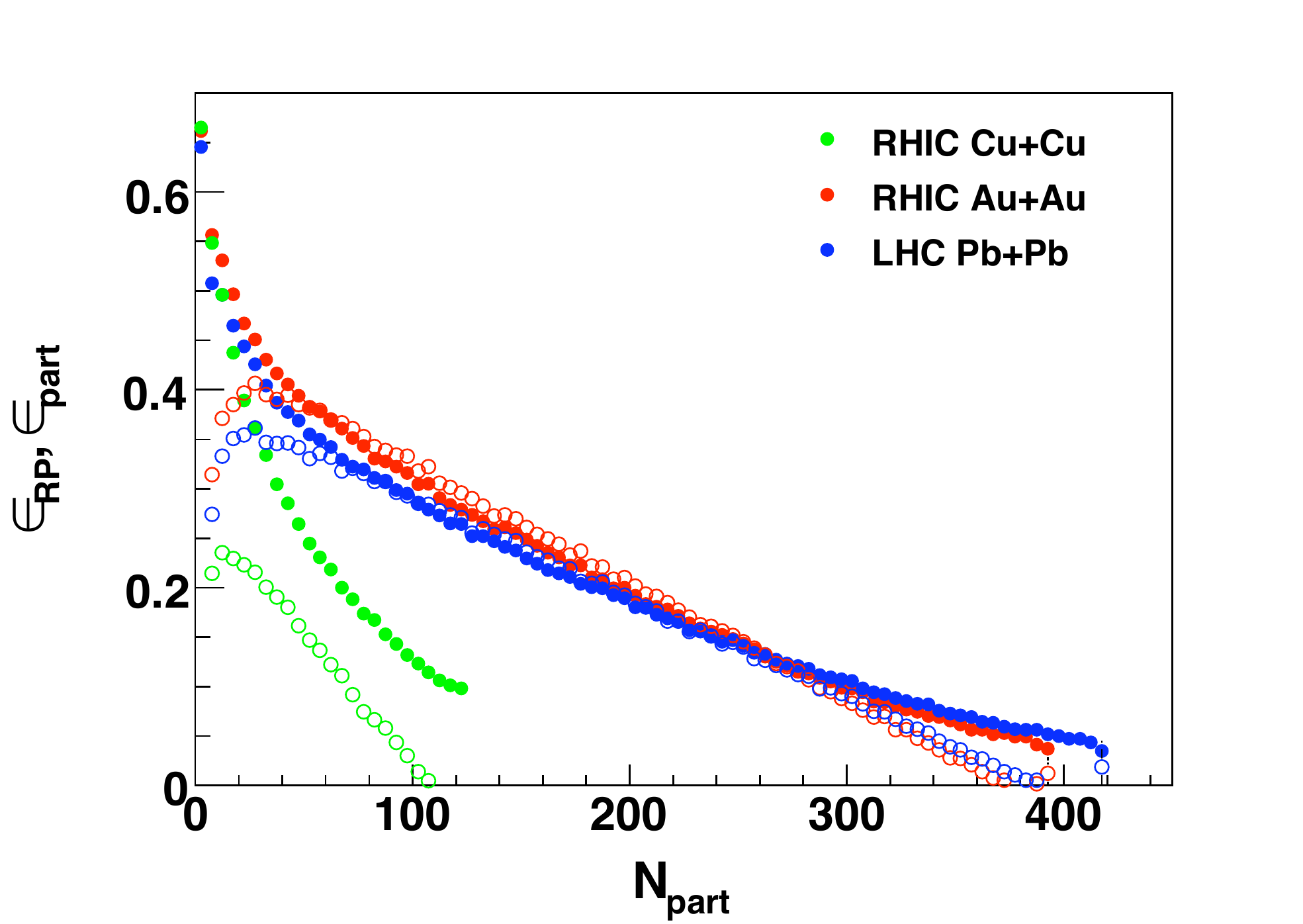}
  \caption{\label{fig:eccplot}$\erp$~(open symbols) and $\ep$~(closed symbols)  
           as a function of $\Npart$ for Cu+Cu and
           Au+Au collisions at RHIC and Pb+Pb collisions at the LHC.}
\end{center}
\end{figure}

\subsection{Sample Results}
As an example application of this code, 10k events were generated for Cu+Cu and Au+Au 
at RHIC energies~($\snn=42$~mb), and Pb+Pb at LHC beam energy~($\snn=72$~mb) using the 
{\tt runAndSaveNtuple()} function. The resulting ntuples were used to plot the distributions 
of $\Npart$ and $\Ncoll$, shown in \Fig{fig:npcplots}. Using the event-by-event
quantities, one can construct combinations of moments like~\cite{eccentricity}:
\begin{itemize}
\item{Reaction-plane eccentricity $\erp$}
\begin{equation}
   \erp = \frac{\text{VarY}-\text{VarX}}{\text{VarY}+\text{VarX}}
\end{equation}
\item{Participant eccentricity $\ep$}
\begin{equation}
   \ep = \frac{\sqrt{(\text{VarY}-\text{VarX})^2+4\text{VarXY}^2}}{\text{VarY}+\text{VarX}}
\end{equation}
\end{itemize}
which are shown in \Fig{fig:eccplot} for the different systems.

\section{Conclusion}
This work has described the PHOBOS implementation of the ``Glauber Model'' 
commonly used by heavy ion physics experiments to study the initial state
configurations of nuclear matter.  The code, accessible online, can be
used within user code or in a standalone mode allowing analysis of 
various distributions (e.g. $N_{part}$, $N_{coll}$, $b$, $\epsilon_{part}$).  
The authors welcome comments on the code and suggestions on how to make
it more useful to both experimentalists and theorists.

Special thanks to Birger Back and Richard Hollis for careful review of the manuscript.
This work was partially supported by U.S. DOE grants
DE-AC02-98CH10886,
DE-FG02-93ER40802,
DE-FG02-94ER40818,  
DE-FG02-99ER41099, and
DE-AC02-06CH11357, by U.S.
NSF grants 9603486, 
0072204,            
and 0245011,        
by Polish KBN grant 1-P03B-062-27(2004-2007),
by NSC of Taiwan Contract NSC 89-2112-M-008-024, and
by Hungarian OTKA grant (F 049823).


\end{document}